\documentclass[a4paper,nofootinbib,superscriptaddress,
secnumarabic,showpacs,,eqsecnum]{revtex4}

\usepackage[T1]{fontenc}

\usepackage{amsmath, amsthm, amsfonts, amssymb, latexsym}
\usepackage{graphicx}
\usepackage{color}
\usepackage{epsfig}

\usepackage[a4paper]{geometry}

\usepackage[caption=false]{subfig}

\usepackage{url}
\usepackage{ifpdf}

\ifpdf
\usepackage[unicode,pdftex]{hyperref}
\hypersetup{
    unicode=true,          
    pdftitle={Energy in shocks},
    pdfauthor={Carlos Herdeiro},
    pdfsubject={General Relativity},
    pdfcreator={LaTeX},          
    pdfproducer={LaTeX},         
    pdfkeywords={General Relativity, Exact solutions,Black Holes},
    colorlinks=false,            
  }
\else
  \usepackage[dvips]{hyperref}
\fi

\def\beq{\begin{equation}}
\def\eeq{\end{equation}}

\def\tu{{\bar u}}
\def\tv{{\bar v}}

\def\trho{{\bar \rho}}

\def\bequ{\begin{equation}}
\def\eequ{\end{equation}}

\begin{document}

\title{\Large Radiation from a $D$-dimensional collision of shock waves: higher order set up and perturbation theory validity}

\author{Fl\'avio S. Coelho, Carlos Herdeiro, Carmen Rebelo and Marco Sampaio}
\affiliation{
Departamento de F\'\i sica da Universidade de Aveiro and I3N \\ 
Campus de Santiago, 3810-183 Aveiro, Portugal}




\date{\today}

\begin{abstract} 
 The collision of two $D$-dimensional, ultra-relativistic particles, described in General Relativity as Aichelberg-Sexl shock waves, is inelastic. In first order perturbation theory, the fraction of the initial centre of mass energy radiated away was recently shown to be $1/2-1/D$. Here, we extend the formalism to higher orders in perturbation theory, and derive a general expression to extract the inelasticity, valid non-perturbatively, based on the Bondi mass loss formula. Then, to clarify why perturbation theory captures relevant physics of a strong field process in this problem, we provide one variation of the problem where the perturbative framework breaks down: the collision of ultra-relativistic \textit{charged} particles. The addition of charge, and the associated repulsive nature of the source, originates an extra radiation burst, which we argue to be an artifact of the perturbative framework, veiling the relevant physics.
 

 \end{abstract}


\maketitle



\section{Introduction}
Recently \cite{Herdeiro:2011ck,Coelho:2012sy}, we have studied the collision of two $D$-dimensional Aichelburg-Sexl (AS) shock waves \cite{Aichelburg:1970dh}, using a method first developed (in $D=4$) by D'Eath and Payne \cite{D'Eath:1992hb,D'Eath:1992hd,D'Eath:1992qu}, with the goal of obtaining the radiated energy. This method is conceptually and technically elaborate, involving both analytical and numerical studies. Remarkably, the fraction of radiated energy - which we refer to as the \textit{inelasticity of the collision}  - agrees in first order perturbation theory, within the numerical error of the method (less than 0.1\%), with the simple formula
\bequ
\epsilon_{\rm 1st \, order}= \frac{1}{2}-\frac{1}{D} \ . \label{miracle}
\eequ
Asymptotically, $\epsilon_{\rm 1st \, order} \rightarrow 1/2$, which agrees with the \textit{bound}, $\epsilon_{AH}$, obtained by computing the apparent horizon (AH) on the past light cone \cite{Eardley:2002re}
 (or on the future one \cite{Yoshino:2005hi}) for a head-on collision of two AS shock waves:
 \begin{equation}
\epsilon_{\rm AH}= 1-\frac{1}{2}\left(\frac{D-2}{2}\frac{\Omega_{D-2}}{\Omega_{D-3}}\right)^{\frac{1}{D-2}} \ , \qquad  \lim_{D\rightarrow\infty}\epsilon_{\rm AH}=\frac{1}{2} \ ,
\end{equation}
where $\Omega_n$ is the volume of the unit $n$-sphere.  Moreover, the trend with $D$ observed from $\epsilon_{\rm 1st \, order}$ agrees, qualitatively, with that of $\epsilon_{\rm AH}$ (cf. Fig. 3 in \cite{Coelho:2012sy}). 

An immediate question is if the appeal and simplicity of the result \eqref{miracle} is kept in higher order perturbation theory. To answer it, the $D$ dimensional formalism developed in \cite{Herdeiro:2011ck,Coelho:2012sy} must be extended beyond linearised theory and, in particular, so does the radiation extraction method. In \cite{Herdeiro:2011ck,Coelho:2012sy}, we have approached the problem of extracting the gravitational radiation by using the Landau-Lifshitz pseudo-tensor, which was straightforward to apply for the setup therein. Indeed, the first order calculation amounts to approximating the outgoing radiation by an isotropic flux, with a value obtained as the limit of the flux at the axis computed in linearised theory. This allowed us to obtain a relatively simple expression from the pseudo-tensor, since only the radiation in the direction of the symmetry axis had to be computed.  In higher order perturbation theory, on the other hand, we will have to include higher order contributions to the news function, 
 following \cite{D'Eath:1992hb,D'Eath:1992hd,D'Eath:1992qu}, which means that
the  pseudo-tensor components become more complicated, making this method less manageable and clear. We shall therefore, in this paper, discuss the higher order perturbative formalism with a different radiation extraction method, valid for generic axially symmetric spacetimes, based on the Bondi mass loss formula in $D$ dimensions~\cite{Tanabe:2011es,Tanabe:2012fg}. In particular, we shall obtain a formula for the inelasticity which is valid to all orders (cf. \eqref{epsilon}), and which makes closer contact with the original Payne and D'Eath computation  \cite{D'Eath:1992hb,D'Eath:1992hd,D'Eath:1992qu}, since it uses the natural generalisation to higher $D$ of the Bondi news function used by these authors. In linearised theory, of course, the result obtained using the Bondi mass loss formula coincides with the pseudo-tensor method.

The second part of this paper is devoted to the study of a collision of shocks obtained from infinitely boosted charged particles. Such shocks were constructed in \cite{Lousto:1988ua,Ortaggio:2006gh} and considered in  \cite{Yoshino:2006dp,Gingrich:2006qh} to estimate an upper bound on the inelasticity, motivated by TeV gravity scenarios \cite{ArkaniHamed:1998rs,Dimopoulos:2001hw, Giddings:2001bu}  (see \cite{Cardoso:2012qm}, Sec. 4, for a recent review). In performing this study, our goal is to probe the validity of a perturbative method to study a process that, at its core, includes a non-perturbative phenomenon - black hole formation. As it turns out, this example illustrates quite well how the method ceases to work, and creates a clear contrast with the neutral case. The reason why the method fails in such a case is that the repulsive nature of the charged gravitational source implies that an important contribution to the radiation, in the perturbative approach, is obtained from the strong field region. In reality, this contribution, or at least an important part of it, should be cloaked by a horizon, as it would be manifest in a non-perturbative computation. Thus, in this example, the perturbative approach entangles the physical signal with a significative spurious signal, making the method uninformative. By contrast, in the neutral case, the spurious signal appears to be subleading.

This paper is organised as follows. In Section \ref{eq:generalframework} we shall set up the higher dimensional perturbation theory for the problem at hand and derive the formula for the inelasticity, leaving some technical details, related to the transformations between the various coordinate systems involved, to Appendix \ref{app:Bondi}. In Section \ref{sec_charge} we discuss the charged case. Althought we have tried to make the discussion self-contained, the construction uses some details provided in  \cite{Herdeiro:2011ck,Coelho:2012sy}. Two technical issues, the gauge fixing to de Donder coordinates and the evaluation of the integral solution for the perturbative metric, were organized as Appendices \ref{gaugetransformation}  and \ref{app:simplify}. We close with some final remarks - Section \ref{sec_final_rem}.




\section{Higher order shock wave perturbation theory}
\label{eq:generalframework}

In~\cite{Herdeiro:2011ck} we have shown that, in a boosted frame, the metric on the future light cone of the collision of two higher dimensional Aichelburg-Sexl shock waves is given by a perturbative series: 
\begin{equation} g_{\mu\nu}=\nu^{\frac{2}{D-3}}\left[\eta_{ \mu \nu}+h_{\mu  \nu}\right]=\nu^{\frac{2}{D-3}}\left[\eta_{ \mu \nu}+\sum_{i=1}^\infty\left(\frac{\lambda}{\nu}\right)^i h_{\mu\nu}^{(i)}\right]\ .\label{eq:pertexpansion} \end{equation}
Here, $\lambda,\nu$ are the energy parameters of the weak/strong shock in the boosted frame, respectively, and the background flat metric in null coordinates is
\begin{equation}
ds^2\equiv \eta_{{\mu}{\nu}}dx^{{\mu}} dx^{{\nu}}  = -2d{u} d{v} + dx^{{i}} dx^{{j}} =-2d{u} d{v}+d\rho^{2} + \rho^2 d {\Omega}^2_{D-3} \ .
\end{equation} 
$d\Omega_{D-3}$ is the line element of the unit $D-3$ sphere. We call this coordinate system \textit{Brinkmann coordinates} \cite{Brinkmann},
where the retarded and advanced times $(\sqrt{2}{u},\sqrt{2}{v})$ are $(t-z,t+z)$ in terms of Minkowski coordinates, and $\{x^{{i}}\}$ are the remaining Cartesian coordinates on the plane of the shocks, ${i}=1\ldots D-2$, such that the transverse radius is ${\rho}=\sqrt{x^{{i}}x_{{i}}}$.  In this Section we will have two further types of coordinates: i) \textit{de Donder} coordinates and ii) \textit{Bondi} coordinates, both to be introduced and explained below. Since the de Donder coordinates coincide with Brinkmann coordinates outside the future light cone of the collision, we adopt the same notation for Brinkmann and de Donder coordinates (as in~\cite{Herdeiro:2011ck}).

The superposition of two shock waves produces boundary conditions on $ u=0$ (the location of the strong shock) which go up to second order. Thus, in second order perturbation theory, the boundary conditions are exact. Suggestively, for $D=4$, the second order result coincides with the outcome of numerical relativity simulations \cite{Sperhake:2008ga}. This sharpens the motivation to pursue this computation (at least) to second order. 

The general perturbative method consists of the following steps. Once we have the boundary data for the perturbation $h_{\mu\nu}|_{{u}=0}$, we insert the ansatz~\eqref{eq:pertexpansion} into the Einstein equations and equate order by order. The components of the metric perturbations do not decouple immediately, but we can perform a gauge transformation to the so called de Donder coordinates so that they indeed decouple. We take the perturbative gauge transformation in the form
\begin{equation}
  x^{ \mu}\rightarrow x^{\mu}+\sum_{i=1}^{+\infty} \left(\frac{\lambda}{\nu}\right)^i\xi^{(i)\mu}(x^{\alpha}) \ ,
\end{equation}
where the vector $\xi^{(i)\mu}(x^{\alpha})$ is to be determined order by order so that the de Donder gauge condition is obeyed; i.e. for $u>0$
\begin{equation}
\label{gauge}
\bar{h}^{(i)\alpha\beta}_{\phantom{(i)\alpha\beta},\beta}=0  \ ,
\end{equation}
(barred quantities are trace reversed). Now $h^{(i)\alpha \beta}$ are the metric perturbations in de Donder coordinates.

Using~\eqref{eq:pertexpansion} and~\eqref{gauge}, the $n$-th order perturbation components obey decoupled wave equations
\begin{equation}\label{eq:nth_order_Feq}
\Box h^{(i)}_{\mu\nu}=T^{(i-1)}_{\mu\nu}\left[h_{\alpha\beta}^{(j<i)}\right] \ ,
\end{equation}
where the source on the right hand side is generated by the lower order perturbations, and can be computed explicitely order by order.

The general integral solution of~\eqref{eq:nth_order_Feq} can be written using the Green's function method (see Theorem~6.3.1 of~\cite{Friedlander:112411})
\begin{equation} \label{eq:sol_orderbyorder}
h^{(i)}_{\mu\nu}=F.P.\int_{u'>0}d^{D}y'\, G(y,y')\left[T^{(i-1)}_{\mu\nu}(y')+2\delta(u')\partial_{v'}h^{(i)}_{\mu\nu}(y')\right] \ ,
\end{equation}
where $F.P.$ denotes the finite part of the integral, $y=\left\{u,v,x_i\right\}$ and we have used the fact that the source only has support in $u>0$. Due to the axial symmetry of the problem, a basis of vectors and tensors on the transverse plane, constructed from $x^i$ and $\delta_{ij}$, is 
\begin{equation}\label{eq:basis_tensors}
\Gamma_i\equiv \dfrac{x_i}{\rho} \; \;, \; \; \; \; \delta_{ij} \; \;, \; \; \; \; \Delta_{ij}\equiv \delta_{ij}-(D-2)\Gamma_i\Gamma_j \;, 
\end{equation} 
where we have chosen the last tensor to be traceless. Then, 
the metric perturbations in de Donder coordinates are decomposed into seven functions of $(u,v,\rho)$, here denoted $A,B,C,E,F,G,H$, in the following way:
\begin{eqnarray}
&h_{uu}\equiv A=A^{(1)}+A^{(2)}+\ldots \qquad &h_{ui}\equiv B \,\Gamma_i =(B^{(1)}+B^{(2)}+\ldots)\Gamma_i  \nonumber\\
&h_{uv}\equiv C=C^{(1)}+C^{(2)}+\ldots \qquad &h_{vi}\equiv F \,\Gamma_i =(F^{(1)}+F^{(2)}+\ldots)\Gamma_i  \label{app:gen_perts}\\
&h_{vv}\equiv G=G^{(1)}+G^{(2)}+\ldots \qquad &h_{ij}\equiv E \,\Delta_{ij}+H\, \delta_{ij} = (E^{(1)}+\ldots) \Delta_{ij}+(H^{(1)}+\ldots) \delta_{ij} \ .\qquad\nonumber 
\end{eqnarray}
With this setup, using the boundary condition on $u=0$ and the solution~\eqref{eq:sol_orderbyorder}, one can find the metric perturbations by solving for these scalars after suitable contractions of~\eqref{eq:sol_orderbyorder} with the tensors~\eqref{eq:basis_tensors}.

\subsection{Extracting the Gravitational Radiation}
\label{sec_ext_gr}

As mentioned in the Introduction, we shall here construct an energy extraction method which is the higher $D$ generalisation of the original method used by D'Eath and Payne, based  on Bondi's news function. The $D$ dimensional extension of the news function formalism has been recently addressed in~\cite{Tanabe:2011es,Tanabe:2012fg}, where the following mass loss formula for the Bondi mass $M_{B}$ was derived: 
\begin{equation}\label{eq:BondiMassLoss}
\dfrac{dM_B}{d\hat\tau}=-\dfrac{1}{32\pi G_D}\int_{S^{D-2}}\dot{\mathfrak{h}}_{\hat I\hat J}^{[1]}\dot{\mathfrak{h}}^{[1]\hat I\hat J}d\Omega_{D-2} \ .
\end{equation}
The dot denotes derivative with respect to the retarded time $\hat \tau$; the $\hat I$ latin indices are raised with the metric components $g_{\hat I\hat J}$; all these quantities, together with the metric functions appearing inside the integral, will be defined in the following. 

The geometry we are considering in de Donder coordinates has the generic form
\begin{equation}\label{eq:metricDeDonder}
ds^2=ds^2_{Flat}+h_{uu}du^2+2h_{ui}dx^{i}du+h_{vv}dv^2+2h_{uv}dudv+2h_{vi}dx^idv+h_{ij} dx^i dx^j \ .
\end{equation}
We can transform it to coordinates $x^{\mu'}=\{\tau,r,\theta,\phi^i\}$, where $r=\sqrt{\rho^2+z^2}$, $\tau=t-r$ and $\theta$ is the angle with the $z$ axis and $\phi^i$ are the angles on the transverse plane, 
and we have used an index with a prime to denote these new coordinates. Then the metric reads:
\begin{equation}\label{eq:metricDeDonderAngular}
ds^2=ds^2_{Flat}+h_{\tau\tau}d\tau^2+2h_{\tau r}d\tau dr+h_{rr}dr^2+2h_{\tau \theta}d\tau d\theta+h_{\theta\theta}d\theta^2+2h_{r\theta}drd\theta+r^2\sin^2\theta h_{\phi\phi}d\Omega_{D-3}^2 \ ,
\end{equation}
where the various components $h_{\mu' \nu'}$ are defined in Appendix~\ref{app:Bondi} in terms of the seven aforementioned scalar functions.

Next, we change to Bondi coordinates $\{x^{\hat \mu}\}$, related to these intermediate coordinates through the transformation
\begin{equation}\label{eq:ChangeBondi}
x^{\mu'}= x^{\mu'}(x^{\hat \alpha}) \ ,
\end{equation}
such that the new $g_{\hat r \hat r}$ and $g_{\hat r \hat \theta}$ metric components vanish, and the form of the metric becomes
\begin{equation}\label{eq:metricBondi}
ds^2=g_{\hat \tau \hat \tau}d\hat\tau^2+2g_{\hat \tau \hat r}d\hat \tau d\hat r+2g_{\hat I \hat \tau}dx^{\hat I}d\hat\tau+g_{\hat I \hat J}dx^{\hat I}dx^{\hat J} \ .
\end{equation}
Here $x^{\hat I}=\left\{\hat\theta,\hat \phi ^i\right\}$ and the radial coordinate $\hat r$ is chosen such that 
\begin{equation}\label{eq:AerialR}
\sqrt{|g_{\hat I\hat J}|}=\hat r^{D-2}\Omega_{D-2}\ .
\end{equation}
Once found such coordinates, assuming that we have an asymptotically flat spacetime with gravitational radiation, it follows that asymptotically~\cite{Tanabe:2011es}
\begin{equation}\label{eq:BondiDecay}
\dfrac{g_{\hat I \hat J}}{\hat r^2}=\Omega_{\hat I \hat J}+\mathfrak{h}_{\hat I\hat J} =\Omega_{\hat I \hat J}+\sum_{k\geq 0}\dfrac{\mathfrak{h}_{\hat I \hat J}^{[k+1]}}{\hat r^{D/2+k-1}} \ ,
\end{equation}
where $\Omega_{\hat I \hat J}$ is the metric on the unit $(D-2)$-sphere and $k$ runs over all integers for $D$ even and over semi-integers for $D$ odd. This defines the $\mathfrak{h}_{\hat I \hat J}^{[1]}$ components appearing in the mass loss formula~\eqref{eq:BondiMassLoss}.

Due to the axial symmetry of our problem one needs not, in fact, transform the angles on the transverse plane, i.e. $\phi^i=\hat\phi^i$, and
\begin{equation}
\mathfrak{h}_{\hat I \hat J}dx^{\hat I} dx^{\hat J}=\mathfrak{h}_{\hat \theta \hat \theta}\,d{\hat\theta}^2+\mathfrak{h}_{\hat \phi \hat \phi}\sin^2\hat\theta\, \Omega_{ij}d\hat\phi^id\hat\phi^j \ .
\end{equation}
From condition~\eqref{eq:AerialR}, we can then eliminate $\mathfrak{h}_{\hat \theta \hat \theta}$ in terms of $\mathfrak{h}_{\hat \phi \hat \phi }$ and find that asymptotically
\begin{equation}
\mathfrak{h}_{\hat \theta \hat \theta}\rightarrow-(D-3)\mathfrak{h}_{\hat \phi \hat \phi}\ ,
\end{equation}
so the mass loss formula can be written as a $\hat \theta$ angular integral of the following angular power flux
\begin{equation}\label{eq:BondiMassSimpler}
\dfrac{dM_B}{d\hat\tau d\cos\hat\theta}=-\dfrac{(D-2)(D-3)\Omega_{D-3}}{32\pi G_D}\lim_{\hat r\rightarrow +\infty}\left[\hat r\hat \rho^{\frac{D-4}{2}} \dot{\mathfrak{h}}_{\hat \phi \hat \phi}\right]^2 \; .
\end{equation}
Using the general form of our metric in de Donder coordinates~\eqref{eq:metricDeDonder}, we have constructed the coordinate transformation~\eqref{eq:ChangeBondi} in Appendix~\ref{app:Bondi}. One then shows that
\begin{equation}\label{eq:BondiMassFinal}
\dfrac{dM_B}{d\hat\tau d\cos\hat\theta}=-\dfrac{(D-2)(D-3)\Omega_{D-3}}{32\pi G_D}\lim_{\hat r\rightarrow +\infty}\left[\hat r\hat \rho^{\frac{D-4}{2}} \left(\dot E+\dot H\right)\right]^2 \ ,
\end{equation}
where all functions are evaluated with Bondi coordinates.  

An important remark is that our derivation does not rely on the metric being perturbative, and therefore it is valid \textit{non-perturbatively}. The assumptions are simply: i) The metric is expressed in de Donder coordinates as in Eq.~\eqref{eq:metricDeDonder}; ii) the metric is axisymmetric, cf. Eq.~\eqref{app:gen_perts}; iii) the spacetime is asymptotically flat and contains  gravitational radiation, cf. Eq.~\eqref{eq:BondiDecay}. 

Furthermore, asymptotically, the Bondi coordinates used in~\eqref{eq:BondiMassSimpler} will approach de Donder coordinates, from the construction in Appendix~\ref{app:Bondi}. So, in general, we can express the mass loss formula in de Donder coordinates as
\bequ
\frac{dM_B}{d{\tau}d\cos \theta}=-\frac{(D-2)(D-3)\Omega_{D-3}}{64\pi G_D}
\lim_{r \rightarrow + \infty} \left[r\rho^\frac{(D-4)}{2}\left(E_{,v} +H_{,v}+E_{,u}+H_{,u}\right) \right]^2\ . \label{eq:bondiadapted}
\eequ
Observe that the $\partial_u$ terms correspond to fluxes across $v={\rm constant}$ surfaces, which are supposed to vanish on $\theta=\pi$, as argued in~\cite{Herdeiro:2011ck}, for the problem of gravitational shock wave collisions (indeed we have checked this numerically). 

We shall now specialize the general formula~\eqref{eq:bondiadapted}, as to facilitate the application of this result to the perturbative problem of shock wave collisions. We assume  spacetime has an ADM energy scale $2\mu$, with which we construct a length scale $L^{D-3}=8\pi G_D \mu/\Omega_{D-3}$. Then, taking units with $L=1$, and dividing by the total ADM energy scale, the inelasticity factor, corresponding to the fraction of radiated energy into gravitational waves, is 
 \begin{equation}
\epsilon_{\rm radiated}= \int_{-1}^{1} \tfrac{d\cos\theta}{2}\lim_{r\rightarrow +\infty}\int d\tau W(\tau,r,\theta)^2\equiv  \int_{-1}^{1} \tfrac{dx}{2}C(x)\label{epsilon} \ ,
\end{equation} 
with
\begin{equation}
W(\tau,r,\theta)\equiv\sqrt{\frac{(D-2)(D-3)}{8}}\, \, r\rho^\frac{D-4}{2}\left(E_{,v} +H_{,v}+E_{,u}+H_{,u}\right) \label{app:waveformDef} \ .
\end{equation}  
$C(x)$ is the generalisation of the Bondi news function (integrated over $\tau$) to higher dimensions. In the appropriate limit, this (more general) result reduces to the one obtained in \cite{Herdeiro:2011ck} using the Landau-Lifschitz method.

In the particular case of the perturbative method for shock wave collisions, the framework is valid for $\theta$ close to $\pi$ ($x \sim -1$). Then, the approximation taken is usually to expand around the axis, and extrapolate off the axis by integrating the truncated expansion over $\theta$. Besides being axially symmetric, our system is invariant under reflections $z\leftrightarrow -z$ so $C(x)$ must be even. Then 
\begin{equation}
\epsilon_{\rm radiated}=  \int_{-1}^{1} \tfrac{dx}{2}\sum_{n=0}^{+\infty}C_n(x^2-1)^{n}=\sum_{n=0}^{+\infty}\dfrac{C_n(-2)^nn!}{(2n+1)!!}\label{epsilonexpandee} \ .
\end{equation} 
If the news function $C(x)$ is analytic, then the expansion close to the axis is indeed sufficient, provided that $\lim_{n\rightarrow +\infty}|C_{n+1}/C_n|\leq 1$. Furthermore, since we have an extra suppression factor in~\eqref{epsilonexpandee}, we expect higher orders to become increasingly less important. The approximation used by D'Eath and Payne~\cite{D'Eath:1992qu} in  $D=4$, corresponds to an isotropy approximation so only $C_0$ is used in their result which give (to second order in perturbation theory) a result of $\epsilon_{\rm radiated}=0.163$. This is in agreement with the latest numerical relativity simulations of ultra-relativistic particle or black hole collisions at large boost~\cite{East:2012mb,Sperhake:2008ga}, so it seems to indicate that the angular corrections ($C_n$ for $n>0$) are small in $D=4$.

\section{$D$ dimensional shock wave collisions with charge}
\label{sec_charge}
\label{section2}

In this Section we shall analyze an example of shock wave collisions with an electric charge parameter. We will apply the formalism developed in the previous Section while testing the assumptions of the formalism and commenting on its limitations. This example will make clear that  the perturbative construction is only applicable if the bulk of the gravitational radiation is generated far away from the strongly curved region of space-time. 

\subsection{The $D$-dimensional metric}
\label{dmetric}

The geometry of the $D$-dimensional Reissner-Nordstr\"om solution with mass $M$ and charge $Q$ is \cite{Tangherlini:1963bw}
\begin{equation}
ds^2 = -V(r)dt^2 + \frac{dr^2}{V(r)} + r^2 d\Omega_{D-2}\ , 
\end{equation}
where
\begin{equation}
V(r)=1- \frac{16\pi G_D M}{ (D-2) \Omega_{D-2}}\frac{1}{r^{D-3}}+\frac{8\pi G_D Q^2}{(D-2)(D-3)} \frac{1}{r^{2(D-3)}}\ .
\end{equation}
It is intuitive, as first argued by Pirani \cite{Pirani}, that the gravitational field of a fast-moving mass should become increasingly similar to that of a gravitational plane-wave, as the speed is increased. For the case of a RN `particle', the corresponding Aichelburg-Sexl  shock wave
is found by boosting this black hole and then taking simultaneously the limit of infinite boost $\gamma$ and vanishing mass and charge, keeping fixed \cite{Lousto:1988ua}

\begin{equation}
\mu=\gamma M \ , \qquad \mathcal{Q}^2=\gamma Q^2 \ .
\end{equation}
 The resulting geometry for a particle moving in the $+z$ direction in Brinkmann coordinates is
\begin{equation}
ds^2 = -2d{u} d{v} + d{\rho}^{2} + {\rho}^2 d {\Omega}^2_{D-3}+\sqrt{2}\kappa \Phi({\rho}) \delta({u}) d{u}^2\ ,\label{AiSe}
\end{equation}
where $\kappa\equiv 8\pi G_D \mu/\Omega_{D-3}$.
The function $\Phi$ depends only on ${\rho}$ and takes the form \cite{Yoshino:2006dp,Gingrich:2006qh}
\begin{equation}
\Phi({\rho},a/\kappa)=-\frac{2a/\kappa}{(2D-7){\rho}^{2D-7}}+\left\{
\begin{array}{ll}
 -2\ln({\rho})\ , &  D=4\  \vspace{2mm}\\
\displaystyle{ \frac{2}{(D-4){\rho}^{D-4}}}\ , & D>4\ \label{phidef}
\end{array} \right. \ ,
\end{equation}
where
\begin{equation}
a\equiv \frac{8\pi^2G_D\mathcal{Q}^2}{D-3}\frac{(2D-5)!!}{(2D-4)!!} \ .
\end{equation}

The above coordinates are discontinuous at the shock. Transforming to Rosen coordinates $(\bar{u},\bar{v},\bar{x}^i)$ \cite{Rosen}, which are continuous at the shock (see \cite{Herdeiro:2011ck} for the explicit transformation), the geometry for two oppositely directed shock waves, with equal charge parameter $a$ in \eqref{phidef}, and equal energy $\mu$ may be written everywhere as a simple superposition of the two individual geometries, except in the future of the collision. Moreover, in a boosted frame, moving with respect to the $({u},{v})$ chart with velocity $\beta$ in the $-z$ direction, the oppositely directed shock waves keep their form, but acquire new energy parameters, respectively,
 \begin{equation}
\kappa \rightarrow e^\alpha \kappa\equiv \nu \ , \qquad \kappa \rightarrow e^{-\alpha} \kappa\equiv \lambda \ ,
\end{equation}
where $e^\alpha=\sqrt{(1+\beta)/(1-\beta)}$. In this boosted frame, the geometry reads
\begin{multline}ds^2 = -2d\tu d\tv + \left[\Big(1+\dfrac{\nu \tu \theta(\tu)}{\sqrt{2}}\Phi''\Big)^2+ \Big(1+\dfrac{\lambda \tv \theta(\tv)}{\sqrt{2}}\Phi''\Big)^2-1\right]d\trho^{2} \\
+ \trho^2\left[\Big(1+ \frac{\nu \tu \, \theta(\tu)}{\sqrt{2} \bar \rho}\Phi'\Big)^2 
 + \Big(1+ \frac{\lambda \tv \, \theta(\tv)}{\sqrt{2} \bar \rho}\Phi'\Big)^2-1\right] d\Omega^2_{D-3}
 \ ,\label{collision}
\end{multline}
which is valid everywhere except in the future light cone of $\tu=\tv=0$.

\subsection{Setting up the perturbative computation and the boundary conditions}
To set up a perturbative computation and derive the geometry in the future light cone of the collision, we proceed as in  \cite{Herdeiro:2011ck}. In the boosted frame one shock carries much more energy than the other and we thus face the weak shock (traveling in the $-z$ direction) as a perturbation of the geometry of the strong shock (traveling in the $+z$ direction). The geometry of the latter is flat for $\bar{u}>0$; thus we make a perturbative expansion of the Einstein equations around flat space-time in this region.
%
%
%
%
%
We shall make the perturbative expansion in Brinkmann coordinates. Moreover, we choose to work with the following rescaled dimensionless coordinates $({u},{v},x^i)\rightarrow\kappa^{1/(D-3)}({u},{v},x^i)$ and rescaled profile function $\Phi({\rho};a/\kappa)\rightarrow\kappa^{-\frac{D-4}{D-3}}\Phi({\rho};a/\kappa^2)$. Thus, hereafter, $\Phi({\rho})\equiv\Phi\left({\rho};a/\kappa^2\right)$ and all coordinates are dimensionless. 

The boundary conditions for the perturbative computation are given by the geometry (\ref{collision}) in the limit ${u}=0^+$, yielding  only the first two orders in~\eqref{eq:pertexpansion}(notice that these boundary conditions are exact, albeit written in a perturbative form):\footnote{The coordinate transformation from Rosen coordinates \eqref{collision} to Brinkmann coordinates is adapted to the strong shock. In particular, the metric remains continuous  at $\bar{v}=0$ after the coordinate transformation. This behaviour at  $\bar{v}=0$ is consistent with the propagation of the initial data from $u=0$, as can be checked numerically.}
\begin{eqnarray}
h^{(1)}_{{u}{u}}&=&-\Phi'^2k({v},{\rho})\ ,\qquad h^{(1)}_{{u}{i}}=\frac{x_{{i}}}{{\rho}}\sqrt{2}\Phi'k({v},{\rho}) \ , \\
h^{(1)}_{{i}{j}}&=&-2\delta_{{i}{j}}h({v},{\rho})-2\frac{x_{{i}}x_{{j}}}{{\rho}^2}\left(k({v},{\rho})-h({v},{\rho})\right) \ ,
\end{eqnarray}
and
\begin{eqnarray}
h^{(2)}_{uu}=\frac{\Phi'^2}{2}k(v,\rho)^2 \ ,\qquad h^{(2)}_{ui}=-\frac{x_i}{\rho}\frac{\Phi'}{\sqrt{2}}k(v,\rho)^2\ , \\
h^{(2)}_{ij}=\delta_{ij}h(v,\rho)^2-\frac{x_ix_j}{\rho^2}\left(k(v,\rho)^2-h(v,\rho)^2\right) \ ,
\end{eqnarray}
where
\begin{equation}
h(v,\rho)\equiv-\frac{\Phi'}{2\rho}(\sqrt{2}v-\Phi)\theta(\sqrt{2}v-\Phi)\ ,\qquad k(v,\rho)\equiv-\frac{\Phi''}{2}(\sqrt{2}v-\Phi)\theta(\sqrt{2}v-\Phi) \ .
\end{equation}

\begin{figure}[t]
\includegraphics[scale=0.2]{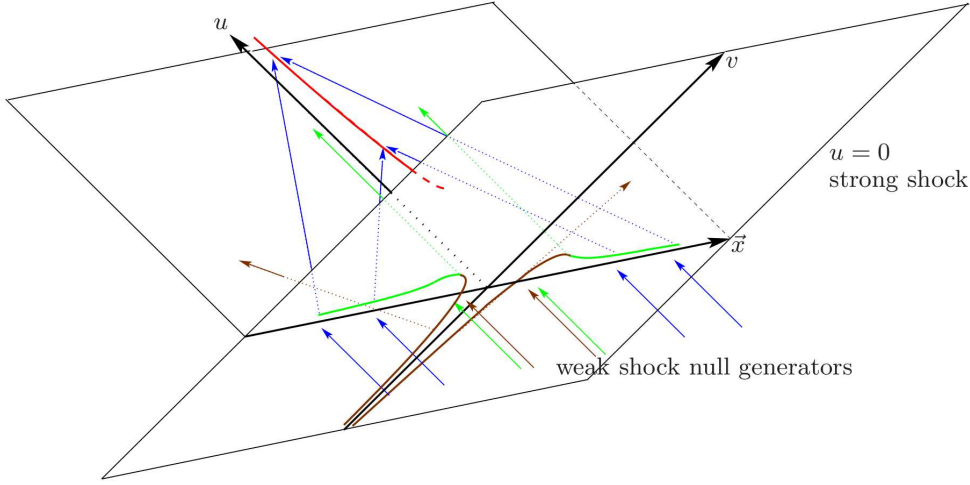}
\caption{Evolution of the weak shock null generators (incident blue, green and brown arrows) from the viewpoint of Brinkmann coordinates in the boosted frame in $D>4$. For $u<0$ they are at $v=0$; then the generators undergo a discontinuity in $v$ at $u=0$, which is $\rho$ dependent and negative for small $\rho$. They jump to the collision surface (green and brown lines). Generically, the rays gain shear and: i)  for large $\rho$ (green part of the curve) focus along the caustic (red line); ii) for small $\rho$ (brown part of the curve) diverge. Undeflected rays are drawn in green.}
\label{lightcone}
\end{figure}

\begin{figure}[t]
\includegraphics[scale=0.25]{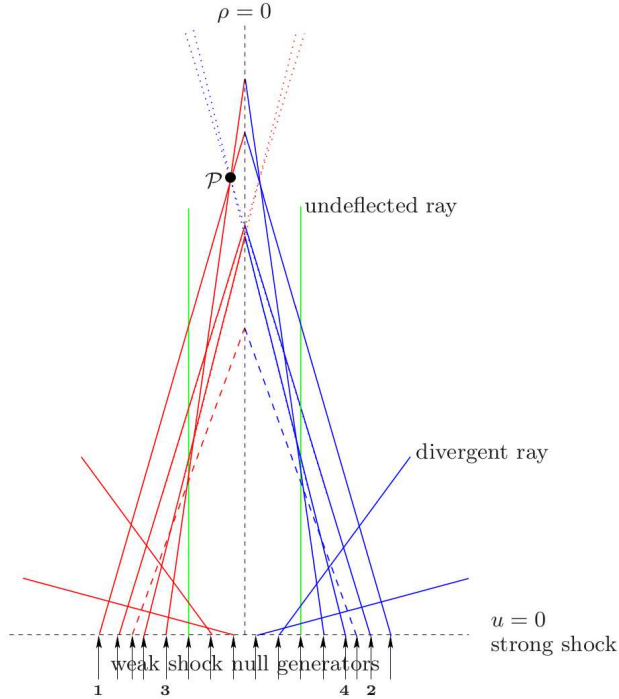}
\caption{Diagram illustrating (a section of) the \textit{spatial} trajectories of the null generators of the weak shock, exhibiting their behaviour after $u=0$. The introduction of charge leads to a divergent behaviour of the rays for $\rho<(a/\kappa^2)^{1/(D-3)}$ and the existence of the undeflected rays (green lines). Moreover, amongst the convergent rays there is one family of maximal deflection (dashed red and blue lines). The convergent rays will meet at the axis forming a caustic. For points outside the axis, far away from the collision, such as point $\mathcal{P}$, the diagram suggests four radiation peaks associated to rays 1-4.  We shall see this interpretation matches the wave forms exhibited in Section \ref{sec_num_rel}. An analogous Figure was presented in \cite{Yoshino:2006dp}.}
\label{rays}
\end{figure}

The step function in the previous equations jumps at the collision, which, in Brinkmann coordinates, occurs at $u=0$, $v= \Phi/\sqrt{2}$. As the weak shock null generators (traveling along $v=0,u<0$) reach the strong shock at $u=0$ there is a discontinuity in Brinkmann coordinates. To understand what occurs, we consider the trajectory of a weak shock null generator that, before the collision, obeys the parametric equations
\bequ
u(\Lambda)=\Lambda \ , \qquad v(\Lambda)=0 \ , \qquad \rho(\Lambda)= \xi \ .
\eequ
Then, after the shock, its trajectory is given by  \cite{Herdeiro:2011ck}
\bequ
u(\Lambda)=\Lambda \ , \qquad 
v(\Lambda) = \theta(\Lambda)\left(\frac{\Phi(\xi)}{\sqrt{2}} + \frac{\Lambda \Phi'(\xi)^2}{4}\right)\ , \qquad \rho(\Lambda)= \xi\left(1-\frac{\sqrt{2}\Lambda \theta(\Lambda)}{\xi^{D-2}}\right) \ .
\eequ
Indeed, the $v$ coordinate jumps from $v=0$ to  the surface $v=\Phi/\sqrt{2}$, i.e. the \textit{collision surface} - Fig. \ref{lightcone}. After this jump the $v$ coordinate of the trajectory increases, \textit{except} at points where $ \Phi'(\xi)=0$. In the uncharged case there were no extrema of this profile function. But in the charged case there is one maximum. The ray incident at the corresponding value of $\xi$ will follow a path of $v={\rm constant}$ after the collision. This is possible for a null trajectory because such ray is \textit{undeflected}. In general the rays are deflected by an angle $\alpha$ in the $u-\rho$ plane, such that
\bequ
\tan \alpha=\frac{\sqrt{2}}{\rho^{D-3}}\left(1-\frac{a/\kappa^2}{\rho^{D-3}}\right) \ .
\eequ
It follows that the weak shock null generators become convergent (towards the symmetry axis) for $\rho>(a/\kappa^2)^{1/(D-3)}$, divergent for $\rho<(a/\kappa^2)^{1/(D-3)}$ and there are undeflected generators for  $\rho=(a/\kappa^2)^{1/(D-3)}$ - Fig. \ref{rays}. This is qualitatively very different from the uncharged case where all rays are convergent (see Fig. 4 in \cite{Herdeiro:2011ck}); the divergent behaviour is caused by the repulsive gravitational effect of the charge.

Finally, for an observation point $\mathcal{P}$ far from the collision and near the axis, Fig. \ref{rays} suggests four bursts of radiation, associated to the four optical paths that reach $\mathcal{P}$. The order in which the four rays should arrive at $\mathcal{P}$ defines their numbering. The first (second) ray comes from the same (opposite) side of the axis as $\mathcal{P}$ and from a low redshift region (cf. Fig. \ref{lightcone}). The third (fourth) ray comes from the same (opposite) side of the axis as $\mathcal{P}$ and from a high redshift region.



\subsection{Gauge Fixing at $u=0$ and Future Development of the Metric}
\label{sec_gauge_fix}
To the future of $u=0$, we use the perturbative expansion introduced in Section~\ref{eq:generalframework} in de Donder gauge. In first order perturbation theory, the gauge fixing condition~\eqref{gauge} does not affect the radiative components in $h^{(1)}_{ij}$;  we refer the interested reader to Appendix~\ref{gaugetransformation} for the details. Using the results in Appendix~\ref{gaugetransformation} we find the initial conditions
\begin{eqnarray}\label{eq:Indpt-Scalars_init1}
E^{(1)}(0,v,\rho)&=&\left(\frac{\Phi'_1}{\rho}+\frac{2D-5}{D-2}\frac{\Phi'_2}{\rho}\right)(\sqrt{2}v-\Phi)\theta(\sqrt{2}v-\Phi) \ , \\ \label{eq:Indpt-Scalars_init2}
H^{(1)}(0,v,\rho)&=&\frac{(D-3)(D-4)}{2(D-2)}\frac{\Phi'_2}{\rho}(\sqrt{2}v-\Phi)\theta(\sqrt{2}v-\Phi) \ ,
\end{eqnarray}
where $\Phi_{1,2}$ are, respectively, the charge (in)dependent parts of $\Phi$. 
To obtain the relevant scalar functions in first order, $E^{(1)},H^{(1)}$, in the future ligth cone of the collision, we use the integral solution~\eqref{eq:sol_orderbyorder}
\begin{equation} \label{eq:sol_orderbyordercharged}
h^{(1)}_{\mu\nu}=F.P.\int_{u'>0}d^{D}y'\, G(y,y')\left[T^{(0)}_{\mu\nu}(y')+2\delta(u')\partial_{v'}h^{(1)}_{\mu\nu}(y')\right] \ .
\end{equation}
The source that must be considered in first order perturbation theory, $T^{(0)}_{\mu\nu}$, is the energy momentum tensor associated to the Maxwell field of the \textit{background} geometry.
By taking the limit described at the beginning of section \ref{dmetric} also for the Maxwell field, one computes that the energy momentum tensor associated to the shock with support at $u=0$ has a single non-vanishing component:
\bequ
T_{uu}=\frac{\mathcal{Q}^2\pi}{\rho^{2D-5}}\frac{(2D-5)!!}{(2D-4)!!}\delta(u) \ .
\label{bsource}
\eequ
First observe that this energy-momentum tensor \textit{does not} have support on $u>0$. Second, this energy-momentum tensor does not source the radiative components of the first order perturbation $h_{ij}^{(1)}$, since only the $T_{uu}$ component is non-vanishing. Finally, this energy-momentum tensor is completely taken into account already by the strong shock geometry, and hence by the initial conditions considered. Indeed, the Einstein tensor computed from (\ref{AiSe}) reads
\bequ
G_{uu}=\frac{(D-3)a}{\rho^{2D-5}}\delta(u) \ ,
\eequ
which, of course, solves the Einstein equations with the source (\ref{bsource}). 

As for the weak shock, it sources an energy momentum tensor with support on $\bar{v}=0$ that, in principle, contributes to the radiative components via the first term in \eqref{eq:sol_orderbyordercharged}. In this paper we shall focus on the second term in  \eqref{eq:sol_orderbyordercharged}, which suffices to demonstrate the difficulties in applying the perturbative method to shock waves with charge.


A summary of the determination of the radiative scalar functions $E^{(1)}, H^{(1)}$,  is described in Appendix~\ref{app:simplify}.

\subsection{Integration Limits}
We can now discuss the domain for the time integration in (\ref{epsilon}). In the uncharged case, we observed in  \cite{Herdeiro:2011ck} that both the beginning of the radiation burst and its peak, as observed at some space-time point $\mathcal{P}$ in the future of the collision, could be understood by a simple ray analysis, similar to that displayed in Figs. \ref{lightcone} and \ref{rays}, together with an analysis of the intersection of the past light-cone of $\mathcal{P}$ with the collision surface. A similar reasoning for the charged case can be made.

Let $\mathcal{P}$ have space-time coordinates $(u,v,\rho)$, or, equivalently $(\tau, r,\theta)$. One now observes that an observation point $\mathcal{P}$, specified by coordinates $(r,\theta)$, close to the axis and far away from the collision is struck by four rays - Fig. \ref{rays}. These arrive at retarded times $\tau_i$, $i=1...4$, where $\tau_i=\tau_i(r,\theta)$ are determined by solving  
\begin{eqnarray}
r\sin\theta=s\bar{\rho}\left(1+\Phi'(\bar{\rho})\frac{\tau+2r\sin^2\frac{\theta}{2}}{2\bar{\rho}}\right)\ ,\\
\tau\left(1-\frac{\Phi'^2(\bar{\rho})}{4}\right)+2r\left(\cos^2\frac{\theta}{2}-\frac{\Phi'^2(\bar{\rho})}{4}\sin^2\frac{\theta}{2}\right)=\Phi(\bar{\rho}) \ , \label{retardedtimes}
\end{eqnarray}
with $s=+1$, for $\tau_1$ and $\tau_3$, and $s=-1$, for  $\tau_2$ and $\tau_4$, simultaneously determining the auxiliary variable $\bar{\rho}$, which now (unlike the uncharged case) has two solutions for each of the two values of $s$, reflecting the two rays coming from each side (cf. Fig. \ref{rays}).

A qualitative difference with respect to the uncharged case is, however, that the past light cone of $\mathcal{P}$ will have a non-vanishing intersection with the collision surface at \textit{all} retarded times, corresponding to points very close to the axis.  We shall now argue, however, that this contribution is unphysical and should be neglected.\footnote{Numerically, this contribution is small but non-zero, in contrast to the neutral case where there is no domain of integration prior to $\tau_1$.} 

In Rosen coordinates (cf. Section \ref{section2}), the collision occurs at $\bar{u}=0=\bar{v}$. In these (continuous) coordinates, the future light cone of the collision has therefore two branches: $\bar{u}=0, \bar{v}>0$ and $\bar{u}>0, \bar{v}=0$. In terms of the Brinkmann coordinates these conditions read:
\bequ 
u=0 \ , \ \  v>\frac{\Phi(\rho)}{\sqrt{2}} \ , \qquad {\rm  and} \qquad u>0 \ , \ \ v=\frac{\Phi(\bar{\rho})}{\sqrt{2}}+u\frac{\Phi'(\bar{\rho})^2}{4} \ , \  \ {\rm where} \ \ \rho=\bar{\rho}\left(1+\frac{u\Phi'(\bar{\rho})}{\sqrt{2}\bar{\rho}}\right) \ . \label{branches} \eequ
The second of these branches determines a surface which intersects the worldline of $\mathcal{P}$ at the retarded times $\tau_i$ computed from (\ref{retardedtimes}). In particular, prior to $\tau_1$, the worldline of $\mathcal{P}$ is \textit{not} in the future light-cone of the collision. Thus, the radiation observed along the worldline of $\mathcal{P}$ before $\tau_1$ is not causally connected to the collision and consequently we neglect it.

%
%

\subsection{Numerical Results}
\label{sec_num_rel}
\begin{figure}[t]
\includegraphics[scale=0.66,clip=true,trim= 0 0 0 0]{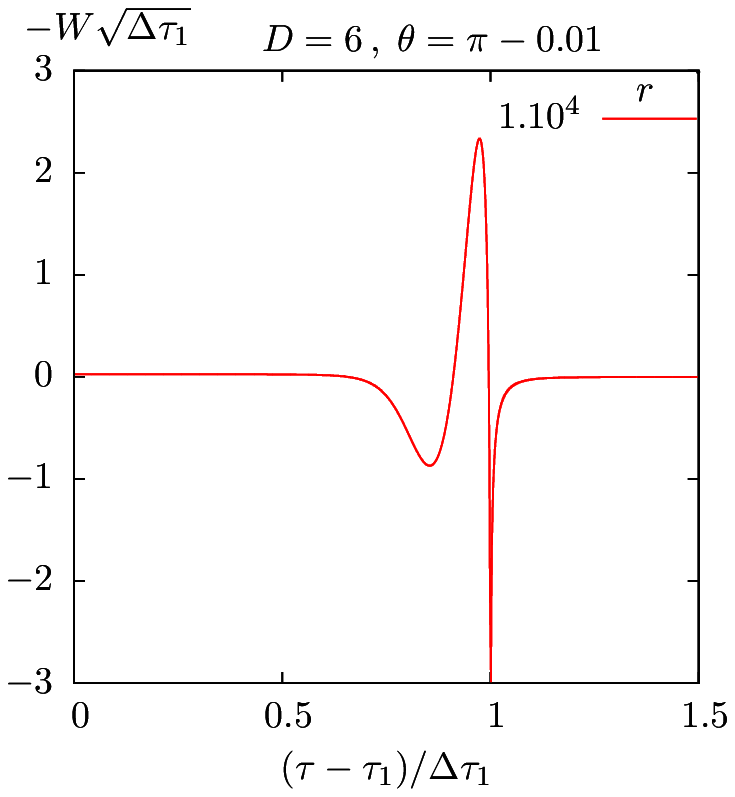} \hspace{1cm} 
\includegraphics[scale=0.66,clip=true,trim= 0 0 0 0]{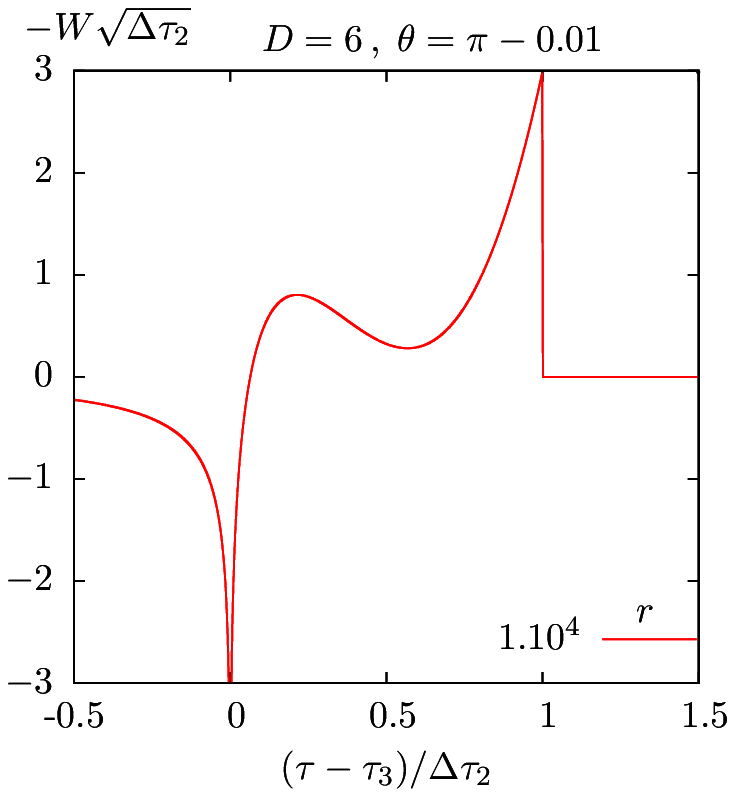} \vspace{3mm}\\
\includegraphics[scale=0.66,clip=true,trim= 0 0 0 0]{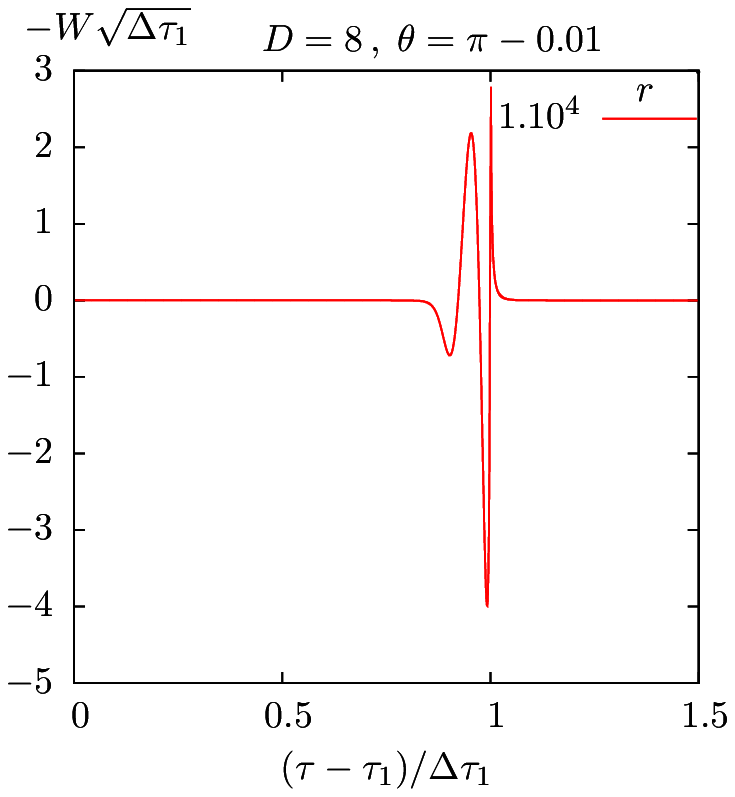}\hspace{1cm}
\includegraphics[scale=0.66,clip=true,trim= 0 0 0 0]{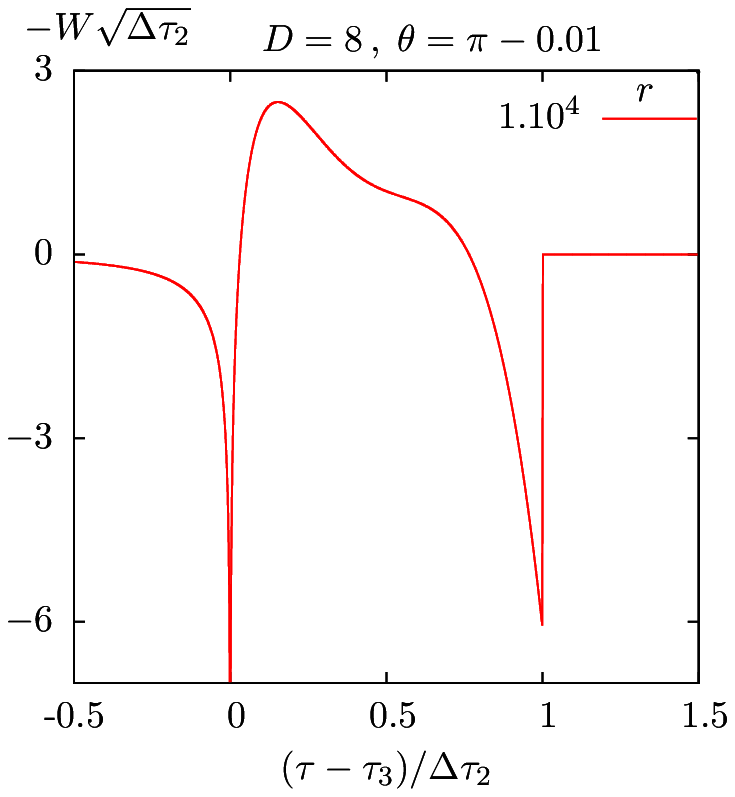} 
\caption{\label{fig:RedWFCharged} {\em Left panels:} First wave form signals with a time scale set by $\Delta \tau_1\equiv \tau_2-\tau_1$, suitably rescaled. {\em Right panels:} Second wave form signals with a time scale set by $\Delta \tau_2\equiv \tau_4-\tau_3$, suitably rescaled. The top panels are for $D=6$ and the bottom ones $D=8$. The signals were generated for $a=0.01,0.1$ and $1$ with no variation in shape.}
\end{figure}

In Fig.~\ref{fig:RedWFCharged}, we display some wave forms for $D=6,8$ obtained from the numerical integration of the contributions in Eqs.~\eqref{wf1}, \eqref{wf2}, \eqref{eq:Ecommau} and~\eqref{eq:Hcommau}. The integration was done using the numerical code developed in~\cite{Herdeiro:2011ck,Coelho:2012sy}. We have found two radiation signals as expected from the geometrical optics analysis. The left hand side plots, corresponding to the signal coming from rays 1 and 2 in Fig. \ref{rays}, coincide precisely with the wave forms computed without charge in~\cite{Herdeiro:2011ck,Coelho:2012sy}. The second signal (right hand side plots), is associated to rays 3 and 4 in Fig. \ref{rays}; these are the rays that are incident close to the axis. In a non-linear computation with horizon formation, a considerable part of this signal should be caught inside the black hole horizon, and should therefore be absent from the viewpoint of an asymptotic observer. The evidence that supports this statement comes from considering the apparent horizon. Indeed, cutting off the $\rho'$ integration region inside the exterior apparent horizon (defined in \cite{Yoshino:2006dp}), the second burst of radiation (between $\tau_3$ and $\tau_4$) vanishes.

Both re-scaled signals are actually independent of the charge parameter for large $r$. This should not be suprising for rays that are incident far away from the center of the collision, since for large $\rho$, the charge contribution to the gravitational field decays faster. However, the second (anomalous) re-scaled signal is also independent of the charge parameter so it gives a constant contribution when integrated in time, to $\epsilon_{\rm radiated}$. In particular this means that the result is discontinuous in the $a\rightarrow 0$ limit, as compared to the $a=0$ result. This is clearly related to the fact that the source is always repulsive for non-vanishing $a$.

We have extracted the two contributions to $\epsilon_{\rm radiated}$ for large $r$, and found that the contribution from the first wave form coincides (within a numerical error of less that 1\%) with the $a=0$ computation. For $D=6$ we found $0.332\pm 0.004$ and for $D=8$, we found $0.374\pm 0.003$. If we add the contribution from the anomalous wave form signal, we get respectively $0.695\pm 0.004$ and $0.876\pm 0.002$, independently of $a$. 

Up to now we have focused on the even $D$ case. For $D$ odd, we expect the method to become even less meaningful since the Green's function for odd $D$ has support not only on the past light cone, but also \textit{inside} the past light cone. Since the shock wave profile becomes repulsive at the center (see Fig.~\ref{lightcone}), we will get contributions to the integrals which come from highly deflected rays that went through the highly curved and non-linear region of spacetime. 

We should also mention that for $D=4$, the integration of the wave form does not even converge for large $r$ to extract a finite $\epsilon_{\rm radiated}$. These results, altogether, indicate the break down of the perturbative method, and clarify its regime of validity. To summarise, the perturbative method should only capture the relevant physics whenever the optical rays arriving at the observation point go through weak field regions only \cite{Herdeiro:2011ck,Coelho:2012sy,D'Eath:1992hb,D'Eath:1992hd,D'Eath:1992qu}.

\section{Final Remarks}
\label{sec_final_rem}

An ultra-relativistic particle collision is a highly dynamical process and also, if a black hole forms as a result, a strong field process. It is then quite remarkable if we may use a perturbative method, such as the one originally proposed in \cite{D'Eath:1992hb,D'Eath:1992hd,D'Eath:1992qu}, to extract relevant physics, such as the inelasticity of the process. 

A good way to test the method is to generalise it. In \cite{Herdeiro:2011ck,Coelho:2012sy} we have extended it to higher dimensions, revealing a remarkably simple pattern in first order perturbation theory. Second order perturbation theory is then the next goal and, in this problem,  one of particular importance. On the one hand,  it will test if the simple pattern observed in \cite{Coelho:2012sy} is a special property of linear theory. On the other hand, the initial data is exact in second order. Furthermore, to this order, an agreement is found (in $D=4$) with numerical relativity simulations. In this paper we have paved the road for this second order computation, by presenting the setup for higher order perturbation theory and formulas for extracting the inelasticity, based on a generalisation of the Bondi mass formula to higher $D$. It remains to compute the relevant scalar functions in second order and to perform the numerical integrations. We shall report on this elsewhere.

Another generalisation considered in this paper was the collision of shock waves with a charge parameter, reminiscent of the Reissner-Nordstr\"om black holes from which they where obtained. These collisions have been considered before for apparent horizon computations  \cite{Yoshino:2006dp,Gingrich:2006qh}, from which bounds on the inelasticity of the process can be obtained. The generic observed behaviour can be described as follows (for simplicity we consider only shock waves with equal charge parameter). Firstly, including the charge parameter  increases the value of $\epsilon_{\rm AH}$, suggesting that a larger fraction of the energy is radiated away. Intuitively, this may be associated to the fact that the charge term in the Reissner-Nordstr\"om (RN) black hole yields a repulsive effect. Secondly, beyond a certain value of the charge parameter no apparent horizon can be seen. Again, this is expected from the RN solution, which has  a limit for the charge to mass ratio, beyond which no event horizon exists. Thirdly, these results are independent of the relative sign of the initial charged particles that were infinitely boosted, in agreement with the observation that gravity is the dominant interaction in trans-Planckian scattering  \cite{'tHooft:1987rb}.

The observed behaviour using the perturbative method to first order shows a qualitative agreement with the first and third observations above. But, as emphasized already, the perturbative method lacks legitimacy in this problem, since there is a considerable amount of radiation that originates in the strong field region, where we have no reason to believe the method. Thus, a more reasonable stance is to regard this charged example as an illustration of how the perturbative method can fail.

%
%

\begin{acknowledgments}
 F.C., M.S. and C.R.  are funded by FCT through the grants SFRH/BD/60272/2009, SFRH/BPD/69971/2010 and SFRH/BPD/77223/2011. The work in this paper is also supported by the FCT grants CERN/FP/116341/2010, PTDC/FIS/098962/2008,  PTDC/FIS/098025/2008, PTDC/FIS/116625/2010, the Marie Curie action NRHEP--295189-FP7-PEOPLE-2011-IRSES, the FCT strategic project PEst-C/CTM/LA0025/2011 and by Funda\c{c}\~{a}o Calouste Gulbenkian through the ``Programa de Est\'imulo \`{a} Investiga\c{c}\~{a}o.''
\end{acknowledgments}

\appendix

\section{Asymptotic construction of Bondi coordinates}
\label{app:Bondi}

In this Appendix we shall address various details related to the coordinate transformations mentioned in Section \ref{eq:generalframework}.  We shall keep the discussion as general as possible, to facilitate its  generalization to higher orders in perturbation theory.

The metric perturbations in the coordinates $\{x^{\mu'}\}$ of Section \ref{eq:generalframework}, are expressed in terms of the scalar functions introduced therein as
\begin{eqnarray}
h_{\tau\tau}&=&\tfrac{A+G}{2}+C \ ,\nonumber\\
h_{\tau r} &=&h_{\tau \tau} + \tfrac{G-A}{2}\cos\theta+\tfrac{B+F}{\sqrt{2}}\sin\theta \ , \nonumber\\
h_{rr}&=&\tfrac{1}{2}\left(A(1-\cos\theta)^2+G(1+\cos\theta)^2\right)+(C+H-(D-3)E)\sin^2\theta+\nonumber \nonumber\\
&&+\sqrt{2}\sin\theta\left(B(1-\cos\theta)+F(1+\cos\theta)\right) \ ,\nonumber\\
h_{\tau \theta}&=&r\left[\tfrac{A-G}{2}\sin\theta+\tfrac{B+F}{2}\cos\theta\right] \ ,\label{eq:MetricTauR}\\
h_{\theta \theta}&=&r^2\left[(\tfrac{A+G}{2}-C )\sin^2\theta+\sqrt{2}(B-F)\sin\theta\cos\theta+(H-(D-3)E)\cos^2\theta\right] \ ,\nonumber\\
h_{r\theta}&=&r\left[\tfrac{1}{2}\left(A(1-\cos\theta)-G(1+\cos\theta)\right)\sin\theta+(C+H-(D-3)E)\sin\theta\cos\theta+\right. \nonumber\\
&&+\left.\tfrac{B+F}{\sqrt{2}}\cos\theta+\tfrac{B-F}{\sqrt{2}}\left(\sin^2\theta-\cos^2\theta\right)\right]\ ,\nonumber\\
h_{\phi\phi}&=&E+H \ .\nonumber
\end{eqnarray}
We consider a generic gauge transformation from de Donder coordinates to Bondi coordinates
\begin{equation}\label{eq:ChangeBondiPert}
  x^{\mu'}=x^{\hat \mu}+\xi^{\hat \mu}(x^{\hat \alpha}) \; ,
\end{equation}
assuming only that $\xi^{\hat \mu}$ decays sufficiently fast with some power of $1/\hat r\sim 1/r$, such that Bondi coordinates match, asymptotically, de Donder coordinates.
Inserting this transformation in the explicit and implicit dependence on $x^{\mu'}$ in equation~\eqref{eq:metricDeDonder} one obtains that the new metric perturbation in Bondi coordinates is asymptotically
\begin{equation}
q_{\hat\mu\hat\nu}=h_{\hat \mu \hat \nu}+\eta_{\hat \mu \hat \nu,\hat \alpha}\xi^{\hat \alpha}+2\eta_{\hat \alpha \left(\hat\nu\right.}\xi^{\hat \alpha}_{,\left.\hat \mu\right)} +\ldots
\end{equation}
where the dots denote terms which decay faster asymptotically, the right hand side terms contain the perturbative metric, and the Minkowski metric evaluated with Bondi coordinates as emphasized by the barred indices. The three conditions to satisfy are
\begin{eqnarray}\label{eq:ConditionsBondi}
q_{\hat r \hat r}&=&q_{\hat \theta \hat r}=0 \ , \\
q_{\hat \theta \hat \theta}&=&-(D-3)q_{\hat \phi \hat \phi} \ ,
\end{eqnarray}
with $q_{\hat \phi \hat \phi}$ defined analogously to $h_{\phi\phi}$. The last condition is only valid, strictly speaking, asymptotically. Using a transformation with $\xi^{\hat \phi_i}=0$, we obtain
\begin{eqnarray}
\xi^{\hat \tau}&=&\int \dfrac{h_{\hat r\hat r}}{2}+\gamma(\hat \tau,\hat \theta) \ , \\
\xi^{\hat \theta}&=&\int \dfrac{1}{\hat r^2}\int \dfrac{h_{\hat r\hat r,\hat\theta}}{2}-\int \dfrac{h_{\hat r \hat \theta}}{\hat r^2}-\dfrac{\gamma(\hat \tau,\hat\theta)_{,\hat \theta}}{\hat r}+\beta(\hat \tau, \hat \theta)\ ,\\
\xi^{\hat r}&=&-\dfrac{\hat r}{2(D-2)}\left[\dfrac{h_{\hat \theta \hat \theta}}{\hat r^2}+(D-3)h_{\hat \phi \hat \phi}-2\left((D-3)\cot \hat \theta \xi^{\hat\theta}+\xi^{\hat\theta}_{,\hat \theta}\right)\right] \ ,
\end{eqnarray}
where $\gamma(\hat \tau,\hat \theta)$ and $\beta(\hat \tau, \hat \theta)$ are two arbitrary integration functions and we have used the symbol $\int$ to denote the primitivation with respect to $\hat r$ (note that the metric functions under the integral decay as an inverse power of $\hat r$). Such integrating functions are further constrained by requiring that the new metric functions in Bondi coordinates decay at large $\hat r$ as in Eq.~\eqref{eq:BondiDecay}. The component we need is finally
\begin{equation}\label{eq:PhiPhi1st}
(D-2)q_{\hat \phi \hat \phi}=h_{\hat \phi \hat \phi}-\dfrac{h_{\hat \theta \hat \theta}}{\hat r^2}+2(\cot \hat \theta-\partial_{\hat \theta})\left(\int \dfrac{1}{\hat r^2}\int \dfrac{h_{\hat r\hat r,\hat\theta}}{2}-\int \dfrac{h_{\hat r \hat \theta}}{\hat r^2}-\dfrac{\gamma_{,\hat \theta}}{\hat r}+\beta\right) \; .
\end{equation}

To have the correct asymptotic decay, the contribution from $\beta$ must be zero; equating the differential operator acting on $\beta $ to zero gives that $\beta=a(\hat \tau)\sin\hat\theta$. The same applies for the $\gamma$ contribution in $D>4$; for $D=4$, however, $\gamma$ remains arbitrary. This is the well known {\em supertranslation} freedom referred to in~\cite{Tanabe:2011es}. Since we are interested in extending the Bondi mass loss formula to higher dimensions, and neither of these arbitrary functions affect the components entering the radiative metric components, we set both contributions to zero in the remainder.

The final quantity that we need is 
$\dot q_{\hat \phi \hat \phi}\rightarrow \mathfrak{\dot h}_{\hat \phi \hat \phi}$,  so all that remains is to take $\hat \tau$ derivatives and use the metric functions in~\eqref{eq:MetricTauR} evaluated in Bondi coordinates. After doing so, the result will still depend on many of the metric functions we have introduced in~\eqref{app:gen_perts}. Such functions are constrained by the de Donder gauge conditions~\eqref{gauge}. Writing down those conditions, changing to $\tau,r,\theta$ coordinates, replacing by hatted coordinates $\hat \tau,\hat r, \hat \theta$, and taking the asymptotic limit one obtains that 
\begin{eqnarray}
\lim_{\hat r\rightarrow+\infty} \hat r^{\frac{D-2}{2}}\left[(1-\cos\hat\theta)\dot A+\sqrt{2}\sin\hat\theta\dot B+\frac{D-2}{2}(1+\cos\theta)\dot H\right]&=&0 \ ,\nonumber\\
\lim_{\hat r\rightarrow+\infty} \hat r^{\frac{D-2}{2}}\left[(1+\cos\hat\theta)\dot G+\sqrt{2}\sin\hat\theta\dot F+\frac{D-2}{2}(1-\cos\theta)\dot H\right]&=&0 \ ,\label{eq:GaugeAsympt}\\
\lim_{\hat r\rightarrow+\infty} \hat r^{\frac{D-2}{2}}\left[(1-\cos\hat\theta)\dot B+(1+\cos\hat\theta)\dot F+\sqrt{2}\sin\hat\theta\left(\dot C-(D-3)\dot E-\tfrac{D-4}{2}\dot H\right)\right]&=&0\nonumber \  . \; \;
\end{eqnarray}
Finally one takes $\hat \tau$ derivatives of~\eqref{eq:MetricTauR} evaluated with hatted coordinates, and inserts relations~\eqref{eq:GaugeAsympt} to obtain
\begin{eqnarray}\label{eq:AsymptoticDeDonder}
\lim_{\hat r\rightarrow+\infty} \hat r^{\frac{D-2}{2}} \dot h_{\hat r \hat r}&=&\lim_{\hat r\rightarrow+\infty} \hat r^{\frac{D-2}{2}} \dot h_{\hat r \hat \theta}=0 \\
\lim_{\hat r\rightarrow+\infty} \hat r^{\frac{D-2}{2}} \dfrac{\dot h_{\hat \theta \hat \theta}}{\hat r^2}&=&-(D-3)\left(\dot E+\dot H\right) \; .
\end{eqnarray}
The final result is then 
\begin{equation}
\dot q_{\hat \phi \hat \phi}\rightarrow \mathfrak{\dot h}_{\hat \phi \hat \phi}=\dot E+\dot H \ ,
\end{equation}
which we have used in \eqref{eq:BondiMassFinal}.

\section{Gauge transformation to the de Donder gauge in the charged case}
\label{gaugetransformation}

In this Appendix we explicitly show that the gauge fixing discussed in Section \ref{sec_gauge_fix} does not affect the radiative components of the metric. For simplicity of notation we shall, herein, omit the superscript ${}^{(1)}$, in the first order metric perturbations. 

Using the integral solution~\eqref{eq:sol_orderbyorder} we can show that the de Donder gauge is preserved in $u>0$  if the following is obeyed on $u=0$: 
\bequ
\bar{h}^{N \beta}_{\phantom{N\beta}\alpha,\beta v}=0 \ ;
\label{DDpreserve}
\eequ
here $N$ is a reminder that this is the new de Donder metric perturbation.
This condition can be written in terms of the boundary values of the perturbations at $u=0$: 
\begin{equation}
\Box\xi_{\alpha,v}=\frac{1}{2}h_{,\alpha v}+h_{\alpha v,uv}+h_{\alpha u,vv}-h_{\alpha i,iv} \ .
\end{equation}
To eliminate the $u-$derivative, we use the Einstein equations
\begin{equation}
h_{\alpha v,uv}=G_{\alpha v}+\frac{1}{2}h_{\alpha v,ii}-\frac{1}{4}\eta_{\alpha v}\Box h+\Box\left[\xi_{(\alpha,v)}-\frac{1}{2}\eta_{\alpha v}\xi^{\gamma}_{,\gamma}\right] \ ,
\end{equation}
to obtain
\begin{equation}
\Box\xi_{[\alpha,v]}+\frac{1}{2}\eta_{\alpha v}\Box\xi^{\gamma}_{,\gamma}=G_{\alpha v}+\frac{1}{2}h_{,\alpha v}-\frac{1}{4}\eta_{\alpha v}\Box h+h_{u\alpha,vv}-h_{i\alpha,iv}+\frac{1}{2}h_{\alpha v,ii}\ .
\label{A3}
\end{equation}
Moreover, the vanishing of the trace of the Einstein tensor implies that $\Box h=-2\Box\xi^{\gamma}_{,\gamma}$; hence (\ref{A3}) becomes
\begin{equation}
\Box\xi_{[\alpha,v]}=G_{\alpha v}+\frac{1}{2}h_{,\alpha v}+h_{u\alpha,vv}-h_{i\alpha,iv}+\frac{1}{2}h_{\alpha v,ii} \ .
\end{equation}
Let us analyse the various components of this equation. For $\alpha=v$, we have an identity
\begin{equation}
0=G_{vv}+\frac{1}{2}h_{,vv} \ ;
\end{equation}
for $\alpha=j$,
\begin{eqnarray}
\Box\xi_{[j,v]}&=&G_{jv}+\frac{1}{2}h_{,jv}+h_{uj,vv}-h_{ij,iv}\nonumber \\
&=&-(2D-5)(D-3)\frac{a}{\kappa^2}\frac{x_j}{\rho^{2D-3}}\sqrt{2}\theta(\sqrt{2}v-\Phi)\ ;
\end{eqnarray}
for $\alpha=u$, we need to eliminate the $u$ derivative of the trace; we can do this by using
\begin{equation}
h_{,uv}=\frac{1}{2}h_{,ii}-\frac{1}{2}\Box h=\frac{1}{2}h_{,ii}+\Box\xi^{\gamma}_{\phantom{\gamma},\gamma}\ ;
\end{equation}
thus $\alpha=u$ reads
\begin{eqnarray}
\Box\left(\xi_{u,v}-\frac{1}{2}\xi_{i,i}\right)&=&G_{uv}+\frac{1}{4}h_{,ii}+h_{uu,vv}-h_{iu,iv}\nonumber \\
&=&\left(F(\rho)(\sqrt{2}v-\Phi)+G(\rho)\right)\theta(\sqrt{2}v-\Phi)\ ,
\end{eqnarray}
where
\begin{eqnarray}
F(\rho)&=&-\frac{a}{2\kappa^2}\frac{1}{\rho^{2D-3}}(2D-5)(D-1)(D-3)\ ,\nonumber \\
G(\rho)&=&(D-3)\left[\frac{4(D-2)}{\rho^{2D-4}}-10(2D-5)\frac{a/\kappa^2}{\rho^{3D-7}}+(19D-51)\frac{a^2/\kappa^4}{\rho^{4D-10}}\right] \ .
\end{eqnarray}

We look for a solution which has a power series expansion around $u=0$:
\begin{equation}
\xi_\mu(u,v,x^i)=\xi_\mu^{(0)}(v,x^i)+u\xi_\mu^{(1)}(v,x^i)+\ldots \ .
\end{equation}
If $\xi_\mu^{(0)}=0$, $\Box\xi_\mu=-2\partial_v\xi_\mu^{(1)}$ (at $u=0$), and we get
\begin{eqnarray}
\xi_{j,v}^{(1)}-\xi_{v,j}^{(1)}&=&(2D-5)(D-3)\frac{a}{\kappa^2}\frac{x_j}{\rho^{2D-3}}(\sqrt{2}v-\Phi)\theta(\sqrt{2}v-\Phi)\ , \\
\xi_{u,v}^{(1)}-\frac{1}{2}\xi_{i,i}^{(1)}&=&-\frac{F(\rho)}{4\sqrt{2}}(\sqrt{2}v-\Phi)^2\theta(\sqrt{2}v-\Phi)-\frac{G(\rho)}{2\sqrt{2}}(\sqrt{2}v-\Phi)\theta(\sqrt{2}v-\Phi)\ .
\end{eqnarray}
One solution is $\xi_v^{(1)}=0$ and:
\begin{eqnarray}
\xi^{(1)}_i&=&\frac{(2D-5)(D-3)a}{2\sqrt{2}\kappa^2}\frac{x_i}{\rho^{2D-5}}(\sqrt{2}v-\Phi)^2\theta(\sqrt{2}v-\Phi)\ , \\
\xi^{(1)}_u&=&-\frac{\bar{F}(\rho)}{24}(\sqrt{2}v-\Phi)^3\theta(\sqrt{2}v-\Phi)-\frac{\bar{G}(\rho)}{8}(\sqrt{2}v-\Phi)^2\theta(\sqrt{2}v-\Phi) \ ,
\end{eqnarray}
where
\begin{eqnarray}
\bar{F}(\rho)&=&F(\rho)+(2D-5)(D-3)(D-1)\frac{a}{\kappa^2\rho^{2D-3}}\\
&=&\frac{a}{2\kappa^2}\frac{1}{\rho^{2D-3}}(2D-5)(D-1)(D-3)\ , \\
\bar{G}(\rho)&=&G(\rho)+(2D-5)(D-3)\frac{a}{\kappa^2}\frac{\Phi'}{\rho^{2D-4}}\ .
\end{eqnarray}
The only metric components that change under this gauge transformation are
\begin{equation}
h^N_{uu}=h_{uu}+2\xi_u^{(1)},\qquad h^N_{ui}=h_{ui}+\xi_i^{(1)} \ ,
\end{equation}
and the trace remains unchanged, $h^N=h$. Thus, the components of the trace-reversed metric perturbation are
\begin{eqnarray}
\bar{h}^N_{uu}=h_{uu}+2\xi_u^{(1)}\ ,\qquad \bar{h}^N_{uv}=h_{uv}+\frac{1}{2}h=\frac{1}{2}h\ ,\qquad \bar{h}^N_{vv}=h_{vv}=0\ , \\
\bar{h}^N_{ui}=h_{ui}+\xi_i^{(1)}\ ,\qquad \bar{h}^N_{vi}=h_{vi}=0,\qquad \bar{h}^N_{ij}=h_{ij}-\frac{1}{2}\delta_{ij}h\ .
\end{eqnarray}
Since only the transverse components will be relevant for the computation of the radiation, we conclude that the coordinate transformation is of no relevance for that matter.

\section{Simplifications of the integral solutions}
\label{app:simplify}
In this Appendix we shall obtain the first order solution for the scalars $E,H$, obeying the initial conditions presented in Section \ref{sec_gauge_fix} for the collision of charged shocks. These are the necessary quantities to compute the inelasticity (cf. Section \ref{sec_ext_gr}). 

Using Eq.~\eqref{eq:sol_orderbyorder} and the definitions of the scalars~\eqref{app:gen_perts}, we obtain explicit integral solutions for the scalars
\bequ
 E^{(1)} = \frac{\sqrt{2}\Omega_{D-4}}{(D-3)(D-1)(2\pi u)^{\frac{D-2}{2}}}\int d\rho'\Psi(\rho')\rho'^{D-4}\int_{-1}^{1}dx\frac{d^2}{dx^2}\left[(1-x^2)^{\frac{D-1}{2}}\right]\delta^{\left(\frac{D-6}{2}\right)}\left(v-v_\star\right) \ ,
\eequ
\bequ
H^{(1)}= \frac{(D-3)(D-4)}{2(D-2)}\frac{\sqrt{2}\Omega_{D-4}}{(2\pi u)^{\frac{D-2}{2}}}\int d\rho'\Phi'_2(\rho')\rho'^{D-4}\int_{-1}^{1} dx (1-x^2)^{\frac{D-5}{2}}\delta^{\left(\frac{D-6}{2}\right)}\left(v-v_\star\right)\  \ ,
\eequ
with 
\begin{equation}
\Psi(\rho')=\Phi'_1(\rho')+\frac{2D-5}{D-2}\Phi'_2(\rho') \ ,\qquad v_\star\equiv\frac{\rho^2-2\rho\rho'x+\rho'^2}{2u}+\frac{\Phi(\rho')}{\sqrt{2}}\ .
\end{equation}
Then, taking into account the scaling properties of the derivatives of the delta function, we obtain
\bequ
E^{(1)}=\frac{\sqrt{2}\Omega_{D-4}}{(D-3)(D-1)(2\pi \rho)^{\frac{D-2}{2}}}\frac{\rho}{u}\int_0^{\infty}d\rho'\Psi(\rho')\rho'^{\frac{D-4}{2}}\int_{-1}^{1}dx\frac{d^2}{dx^2}\left[(1-x^2)^{\frac{D-1}{2}}\right]\delta^{\left(\frac{D-6}{2}\right)}(x-x_{\star})\ , 
\eequ
\bequ
H^{(1)}=\frac{(D-3)(D-4)}{2(D-2)}\frac{\sqrt{2}\Omega_{D-4}}{(2\pi\rho)^{\frac{D-2}{2}}}\frac{\rho}{u}\int_0^{\infty} d\rho'\Phi'_2(\rho')\rho'^{\frac{D-4}{2}}\int_{-1}^{1} dx (1-x^2)^{\frac{D-5}{2}}\delta^{\left(\frac{D-6}{2}\right)}\left(x-x_\star\right) \ ,
\eequ
where
\begin{eqnarray}
x_\star&\equiv&\frac{U\Phi(\rho')+\rho'^2-UT}{2\rho\rho'}\ , \qquad U\equiv\sqrt{2}u \ , \qquad T\equiv\sqrt{2}v-\rho^2/U \ .
\end{eqnarray}

Similarly, after taking a $v$ derivative and using the scaling properties of the derivative of the delta distribution, we obtain
\bequ
E^{(1)}_{,v}=\frac{\sqrt{2}\Omega_{D-4}}{(D-3)(D-1)(2\pi \rho)^{\frac{D-2}{2}}}\int_0^{\infty}d\rho'\Psi(\rho')\rho'^{\frac{D-6}{2}}\int_{-1}^{1}dx\frac{d^2}{dx^2}\left[(1-x^2)^{\frac{D-1}{2}}\right]\delta^{\left(\frac{D-4}{2}\right)}(x-x_{\star})\ , \label{wf1}
\eequ
\bequ
H^{(1)}_{,v}=\frac{(D-3)(D-4)}{2(D-2)}\frac{\sqrt{2}\Omega_{D-4}}{(2\pi\rho)^{\frac{D-2}{2}}}\int_0^{\infty} d\rho'\Phi'_2(\rho')\rho'^{\frac{D-6}{2}}\int_{-1}^{1} dx (1-x^2)^{\frac{D-5}{2}}\delta^{\left(\frac{D-4}{2}\right)}\left(x-x_\star\right) \ . \label{wf2}
\eequ
Finally, 
\begin{eqnarray}
E^{(1)}_{,u}&=&-\frac{D-2}{2u}E^{(1)}-\frac{\sqrt{2}\Omega_{D-4}}{(D-3)(D-1)(2\pi \rho)^{\frac{D-2}{2}}}\times \nonumber\\
&&\times\int_0^{\infty}d\rho'\Psi(\rho')\rho'^{\frac{D-6}{2}}\int_{-1}^{1}dx\frac{d^2}{dx^2}\left[(1-x^2)^{\frac{D-1}{2}}\right]\frac{\rho^2-2\rho\rho'x+\rho'^2}{2u^2}\delta^{\left(\frac{D-4}{2}\right)}(x-x_{\star})\ , \; \; \; \; \label{eq:Ecommau}\\
H^{(1)}_{,u}&=&-\frac{D-2}{2u}H^{(1)}-\frac{(D-3)(D-4)}{2(D-2)}\frac{\sqrt{2}\Omega_{D-4}}{(2\pi\rho)^{\frac{D-2}{2}}}\times\nonumber\\
&&\times\int_0^{\infty} d\rho'\Phi'_2(\rho')\rho'^{\frac{D-6}{2}}\int_{-1}^{1} dx (1-x^2)^{\frac{D-5}{2}}\frac{\rho^2-2\rho\rho'x+\rho'^2}{2u^2}\delta^{\left(\frac{D-4}{2}\right)}\left(x-x_\star\right)\ . \label{eq:Hcommau}
\end{eqnarray}

One important point resulting from the analysis performed so far is that, as for the uncharged case, the relevant functions for the radiation extraction (\ref{wf1}) and (\ref{wf2}) have an argument with a (fractional) derivative of a delta function. The vanishing of the argument of the delta function has the same interpretation as in the uncharged case, namely the intersection of the past light cone of the observation point with the collision curve.

To deal with the $x$ integration it is convenient to separate even and odd $D$. In the following we shall only consider the even $D$ case, since it suffices to illustrate the problems with the method.

\subsection{$D$ even integrals}

\begin{eqnarray}
\int_{-1}^{1}dx\frac{d^2}{dx^2}\left[(1-x^2)^{\frac{D-1}{2}}\right]\delta^{\left(\frac{D-6}{2}\right)}(x-x_{\star})&=&(-1)^\frac{D-2}{2}\dfrac{d^\frac{D-2}{2}}{dx_\star^\frac{D-2}{2}}\left[\left(1-x_\star^2\right)^\frac{D-1}{2}\right]\theta(1-x_\star)\theta(1+x_\star) \nonumber\\
&\equiv& (-1)^\frac{D-2}{2}\sqrt{1-x_\star^2}\sum_{n=0}^{\frac{D-2}{2}}a_n x_\star^n\;\theta(1-x_\star)\theta(1+x_\star)\nonumber \ ;\\
\end{eqnarray}
\begin{multline}
\int_{-1}^{1} dx(1-x^2)^{\frac{D-5}{2}}\delta^{\left(\frac{D-6}{2}\right)}\left(x-x_\star\right)\nonumber\\= \left\{\begin{array}{ll} \left[\frac{\pi}{2}-\arcsin x_\star\right]\theta(1-x_\star)\theta(1+x_\star)+\pi\theta(-1-x_\star)& , \ D=4\ , \\&\\ (-1)^\frac{D-2}{2}\dfrac{d^\frac{D-6}{2}}{dx_\star^\frac{D-6}{2}}\left[\left(1-x_\star^2\right)^\frac{D-5}{2}\right]\theta(1-x_\star)\theta(1+x_\star)&  , \ D\geq 6 \ ,\end{array}\right.\nonumber
\end{multline}
with
\begin{equation}
\dfrac{d^\frac{D-6}{2}}{dx_\star^\frac{D-6}{2}}\left[\left(1-x_\star^2\right)^\frac{D-5}{2}\right]\equiv \sqrt{1-x_\star^2}\sum_{n=0}^\frac{D-6}{2}b_n x_\star^n \ ;
\end{equation}
\begin{eqnarray}
\int_{-1}^{1}dx\frac{d^2}{dx^2}\left[(1-x^2)^{\frac{D-1}{2}}\right]\delta^{\left(\frac{D-4}{2}\right)}(x-x_{\star})&=&(-1)^\frac{D}{2}\dfrac{d^\frac{D}{2}}{dx_\star^\frac{D}{2}}\left[\left(1-x_\star^2\right)^\frac{D-1}{2}\right]\theta(1-x_\star)\theta(1+x_\star)\nonumber\\
&\equiv&(-1)^\frac{D}{2}\dfrac{1}{\sqrt{1-x_\star^2}}\sum_{n=0}^\frac{D}{2}c_n x_\star^n\,\theta(1-x_\star)\theta(1+x_\star) \ ;\nonumber\\
\end{eqnarray}
\begin{eqnarray}
\int_{-1}^{1} dx (1-x^2)^{\frac{D-5}{2}}\delta^{\left(\frac{D-4}{2}\right)}\left(x-x_\star\right)&=&(-1)^\frac{D-4}{2}\dfrac{d^\frac{D-4}{2}}{dx_\star^\frac{D-4}{2}}\left[\left(1-x_\star^2\right)^\frac{D-5}{2}\right]\theta(1-x_\star)\theta(1+x_\star)\nonumber\\
&\equiv&(-1)^\frac{D-4}{2}\dfrac{1}{\sqrt{1-x_\star^2}}\sum_{n=0}^\frac{D-4}{2}d_n x_\star^n\,\theta(1-x_\star)\theta(1+x_\star) \ ;
\end{eqnarray}
\begin{eqnarray}
&&\int_{-1}^{1}dx\frac{d^2}{dx^2}\left[(1-x^2)^{\frac{D-1}{2}}\right]\frac{\rho^2-2\rho\rho'x+\rho'^2}{2u^2}\delta^{\left(\frac{D-4}{2}\right)}(x-x_{\star})\nonumber\\
&&\hspace{2cm}=(-1)^\frac{D-4}{2}\dfrac{d^\frac{D-4}{2}}{dx_\star^\frac{D-4}{2}}\left[\dfrac{d^2}{dx_\star^2}\left(\left(1-x_\star^2\right)^\frac{D-1}{2}\right)\frac{\rho^2-2\rho\rho'x_\star+\rho'^2}{2u^2}\right]\theta(1-x_\star)\theta(1+x_\star) \nonumber\\
&&\hspace{2cm}\equiv (-1)^\frac{D-4}{2}\dfrac{1}{\sqrt{1-x_\star^2}}\sum_{n=0}^\frac{D+2}{2}g_n x_\star^n\,\theta(1-x_\star)\theta(1+x_\star) \ ;
\end{eqnarray}
\begin{eqnarray}
&&\int_{-1}^{1} dx (1-x^2)^{\frac{D-5}{2}}\frac{\rho^2-2\rho\rho'x+\rho'^2}{2u^2}\delta^{\left(\frac{D-4}{2}\right)}\left(x-x_\star\right)\nonumber\\
&&\hspace{1cm}=(-1)^\frac{D-4}{2}\dfrac{d^\frac{D-4}{2}}{dx_\star^\frac{D-4}{2}}\left[\left(1-x_\star^2\right)^\frac{D-5}{2}\frac{\rho^2-2\rho\rho'x_\star+\rho'^2}{2u^2}\right]\theta(1-x_\star)\theta(1+x_\star) \nonumber\\
&&\hspace{1cm}\equiv (-1)^\frac{D-4}{2}\dfrac{1}{\sqrt{1-x_\star^2}}\sum_{n=0}^\frac{D-2}{2}h_n x_\star^n\,\theta(1-x_\star)\theta(1+x_\star) \ .
\end{eqnarray}

\end{document}